\documentclass[conference]{IEEEtran}
\IEEEoverridecommandlockouts
\usepackage{cite}
\usepackage{amsmath,amssymb,amsfonts}
\usepackage{algorithmic}
\usepackage{graphicx}
\usepackage{textcomp}
\usepackage{xcolor}
\usepackage{booktabs}
\usepackage{setspace,url}
\usepackage{subcaption}
\renewcommand{\baselinestretch}{1.05} % Adjust to 1.1 or 1.2 if needed

\def\BibTeX{{\rm B\kern-.05em{\sc i\kern-.025em b}\kern-.08em
    T\kern-.1667em\lower.7ex\hbox{E}\kern-.125emX}}
\begin{document}

\title{Can LLMs Resolve Entity Conflicts in Domain Specific NLU Pipelines?\\
\thanks{}
}

\author{
    \IEEEauthorblockN{
        % SDS 2025 Anonymous Submission
        Akansha Bhardwaj     \\
        Mingbo Cui\\
        Victor Guimarães\\
        Dominik Kohler\\
        Diana Nicoleta Popa\\
        Meghana M. Reddy\\
        Paolo Rosso \\ \\ 
    }
    \IEEEauthorblockA{
        Telepathy Labs GmbH, Zürich, Switzerland\\
        Email: {firstname.lastname}@telepathy.ai
    }
}
\maketitle

\begin{abstract}
Resolving entity conflicts is a critical step in Pipeline-based NLU architectures, especially when multiple extractors provide overlapping or conflicting outputs. Traditional rule-based entity conflict resolution methods often lack flexibility and are unscalable. In this work, we aim to study if the broad knowledge of LLMs can be leveraged to resolve entity conflicts in domain specific NLU pipelines. We present an experimental comparison between a traditional rule-based conflict resolution method, and two LLM-based approaches: a multi-shot prompting method using the~\texttt{Llama3.1-8B-Instruct} model and a LoRA fine-tuned~\texttt{Llama3.1-8B-Instruct} model. The evaluation was conducted under two setups: (1) a broad assessment of LLM's capability to match existing NLU pipeline output, and (2) a more controlled scenario assessing the LLM accuracy on human-labeled data. Results show that while multi-shot prompting offers ease of adaptation and rapid behavioral changes, the fine-tuned model delivers superior performance on complex, ambiguous cases. Ultimately, the best approach appears to be a combination of both strategies, leveraging multi-shot prompting's flexibility alongside the broad accuracy gains of LoRA fine-tuning for a more comprehensive performance.
\end{abstract}


\section{Introduction}

Named Entity Recognition~(NER) is a fundamental task in Natural Language Understanding~(NLU) pipelines which aim to understand information from text~\cite{vYadav18}. While some pipelines use a single NER extractor to handle various entity types, domain-specific applications commonly apply multiple NER extractors and leverage the expertise of each extractor~\cite{tBocklisch17}. This is also a modular approach to entity extraction. For instance, if a new service package \emph{toyota care}, has to be recognized, it can be integrated without impacting other NER extractors in the pipeline.

\begin{figure*}
\begin{minipage}[t]{1.0\linewidth}
 \begin{center}
  \includegraphics[width=0.8\textwidth]{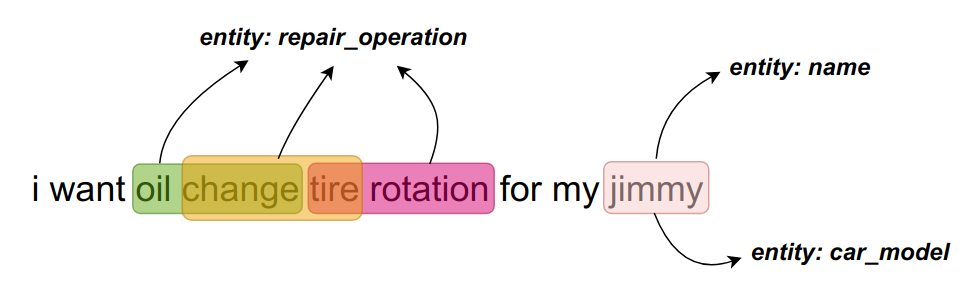}
   \end{center}
\end{minipage}
   \caption{An example of overlapping and conflicting entity spans from automotive domain. A rule-based extractor can incorrectly extract \emph{change tire} as a repair operation from the utterance \lq\emph{oil change tire rotation}' where the caller intended repair operations are \emph{oil change} and \emph{tire rotation}. To resolve this conflict, a rule-based conflict resolution strategy leverages the span information. Similarly, a data driven NER incorrectly extracts \emph{jimmy} as a \emph{name} entity type, but \emph{jimmy} is a car model from GMC, a division of American automotive manufacturer General Motors. To resolve this conflict, a rule-based conflict resolution strategy leverages the intent of the utterance and the presence of other entities. Such rule-based conflict resolution strategies are specific to each use-case and tedious to maintain.}
  \label{fig:overlapping_and_conflicting_entity_span_example}
\end{figure*}

When applying multiple NER extractors, predictions of extractors might conflict with each other by assigning different entity types to overlapping spans of text. Figure \ref{fig:overlapping_and_conflicting_entity_span_example} shows an example from automotive domain where the extracted entities \emph{oil change}, \emph{change tire} and \emph{tire rotation} of \textit{repair\_operation} category overlap with one another. Additionally, \emph{jimmy} is recognized as both \textit{car\_model} and \textit{name} entities. However, the correct set of entities in this context are \emph{oil change}, \emph{tire rotation}, and \emph{jimmy}~(recognized as a \textit{car\_model}).

NLU systems like spaCy \cite{mHonnibal15} usually address this problem by applying the NER extractors sequentially, such that NER extractors later in the pipeline have a lower priority~(to assign entities in case of span overlap) over extractors earlier in the pipeline. This pipeline-based approach relies on a curated order of the NER extractors. However, determining the perfect order of these extractors is challenging. In our use case of a conversation agent, the perfect order of extractors would also depend on the state of the dialogue, which makes this approach infeasible for our needs.

An alternative approach that overcomes the limitations of ordered NER extractors consists in applying multiple NER extractors in parallel and collecting all predicted entities. Naturally, an additional step is to resolve conflicting entities. One approach to entity conflict resolution is to write hand-crafted rules that define prioritization of entities given the current state of the dialogue.
% In order to overcome this limitation, we currently apply multiple NER extractors in parallel, and collect the entities predicted by all extractors. Then, we apply manually-crafted rules to decide which entities to keep and which entities to discard, given the current state of the dialogue.
However, this approach comes with its own limitations -- designing these rules requires domain expertise; and adding new rules is challenging as they might interfere with existing rules.

Recently, Large Language Models (LLMs) have been a recurring topic in the NLU community, given their ability to deal with different NLU tasks with great success~\cite{wZhao23}. In this paper, we investigate the effect of leveraging LLMs to replace the set of manually-crafted rules for entity conflict resolution. We first generate a dataset containing over $60000$ utterance examples from our real-world dialogue system. We then use this dataset to train and evaluate LLMs and compare their performance against our rule-based conflict resolution approach in multiple settings. Our experiments show that LLM-based conflict resolution approach achieves an agreement of $81.42\%$ with the rule-based conflict resolution approach, indicating that LLMs are valid alternatives to consider as replacements for tedious hand-crafted rule-based conflict resolution strategies in specific domains.

The rule-based conflict resolution approach, however, is not a perfect standard to evaluate the performance of LLMs as it might fail to capture more nuanced aspects of the text. To objectively evaluate the effectiveness of our proposed approach, we create a human-annotated dataset, \emph{Automotive-Curated} that contains $674$ examples from three distinct conversation contexts. Results show that on this dataset, the LLM and rule-based conflict resolution approach achieved a competitive overall accuracy of $78.93\%$, and $78.60\%$ respectively. Further, we observe that LLM outperforms rule-based conflict resolution approach in generic conversational context like \emph{time negotiation}, but the comparative performance drops in domain specific contexts like \emph{service request}.

\section{Related Work}

A well-known pipeline based NLU architecture is spaCy \cite{mHonnibal15}, where multiple entity extractors can be included in the pipeline to handle various types of entity extraction tasks. Such architectures commonly apply multiple NER extractors sequentially to a given text. In case of conflicts, the order of extractors in the pipeline configuration defines the priority of extracted entities, where entities extracted through subsequent extractors are de-prioritized over entities extracted through extractors defined earlier in the pipeline. Another common conflict resolution strategy is a custom component that handles resolution of specific conflicts, instead of prioritizing based on the order of NER extractor.

Recently, NER approaches have shifted toward span-based extractors~\cite{jFu21}. These kind of extractors work by first identifying potential spans in the text, then classifying the spans as relevant entities or not. Usually, a final step to resolve overlapping entities is required. Several approaches have been proposed to address this problem -- one involves using Semi-Markov conditional random fields to learn the probabilities of label sequences and selecting the sequence with the highest probability~\cite{sSarawagi2004}, while another uses a greedy method that retains the prediction with the highest score and discards overlapping ones~\cite{yKi21}. Different aspects of span-based NERs are studied in~\cite{uZaratiana22b}. None of these approaches are suitable for our use case as we combine different NER extractors, some of which are black-boxes, and we do not have access to the prediction probabilities. % Since we apply different extractors, where some are black-boxes, and we do not have access to the probability of the predictions, theses approaches are not suitable for uses. %% Furthermore, the predictions would not be comparable among different extractors.

Other methods such as \cite{xHu24,uZaratiana22a,kYu20} learn a unifying extractor to combine the output of the different NER extractors, and thus, are closest related to our approach. However, these methods usually rely on implementation details of the combined NER extractors and/or the construction of additional data structures like a graph relation between predicted entities~\cite{uZaratiana22a}, or a hierarchy among the possible tags~\cite{kYu20}.

% On the other hand, our proposed approach only requires the predictions of the NER extractors, treating them as black-boxes; the original text; and an additional string to indicate from which part of the dialogue the text was extracted.

On the contrary, our approach makes no assumptions on the underlying NER extractor design, making it fit for integrating multiple and potentially different NER extractors, even when they function as black-boxes. The only inputs required by our approach are -- the original text, predictions from the NER extractors, and an additional string indicating the context of the dialog from which the text was extracted.
% Also, maybe this idea can be re-used in the introduction or motivation somewhere to outline the "NER-extractors agnostic compatibility of our approach

\section{The LLM-based Conflict Resolution Component}

The LLM-based conflict resolution component is a sub task in the NLU pipeline that leverages a LLM to handle all entity-related conflict resolution tasks. Instead of relying on multiple rule-based entity conflict resolution components for different entity types or domains, this single component can resolve entity conflicts across various contexts, leading to a simplified and more maintainable pipeline architecture.

\subsection{Motivation}

The existing NLU pipeline contains different NER components that are able to extract multiple entities related to date, time, person names, numerical values, and service requests. To resolve entity conflicts, such a pipeline requires multiple distinct conflict resolution components.

Each of these components has to be carefully configured and maintained, since a change in one component can affect the other components adversely. This further increases the complexity of the pipeline. The LLM-based conflict resolution component is designed to unify these tasks into a single step, improving efficiency and reducing the need for individual specialized components.

\subsection{Implementation Details}

The LLM-based component operates by leveraging the contextual understanding and generalized knowledge of the underlying LLM. The component receives a list of extracted entities from multiple extractors and uses the LLM to disambiguate them based on the surrounding context. Here’s how the approach works in practice:
\newline

\begin{enumerate}

    \item The user’s input is processed by multiple NER extractors in the NLU pipeline, which identify a variety of entities.
    
    \item The LLM-based component receives these entities along with the input message and conversation context.
    
    \item The component then uses the LLM to resolve ambiguities, taking into account the context of the conversation, and domain-specific knowledge.
    
    \item The LLM predicts the most relevant entities or the correct interpretation, and resolves conflicts across different entity types (e.g., distinguishing between \emph{1500} in the utterance \lq\emph{1500 ram}' as a \textit{car\_model} or a \textit{number} entity).
    
    \item At this stage, the resolved entities are passed downstream of NLU, to be used in subsequent dialogue management or action execution steps. If there are overlapping entities in the output at this stage, the dialog management component prioritizes entities based on its policies.

\end{enumerate}

\subsection{Advantages}

The LLM-based disambiguation component offers several key advantages over the previous multi-component approach:
\newline
\begin{itemize}
    \item \textbf{Simplified and model-agnostic Pipeline:} By replacing multiple conflict resolution components with a single LLM-based solution, the pipeline becomes easier to maintain. Previously, each conflict resolution component needed to be manually configured for specific domains. The LLM-based approach reduces this configuration effort by generalizing the disambiguation process.

    \item \textbf{Improved Accuracy:} LLMs can potentially capture more nuanced relationships between entities and context, leading to higher accuracy in disambiguating entities, especially in complex or ambiguous sentences. For instance, identifying if \emph{none} should map to a \textit{number} entity with value $0$, or is irrelevant in the sentence \lq\emph{none of your business, transfer me to customer care}'.

    \item \textbf{Context-Aware Resolution:} The component leverages the LLM’s ability to use conversation context and domain knowledge, which allows for more intelligent and context-aware resolution of conflicting entities. For example, a \emph{number} entity in \emph{contact details} conversational context is probably a user's phone number rather than the vehicle's mileage.
\end{itemize}

\section{Experimental Setup}

In this section, we describe our training and evaluation datasets, experimental setup and evaluation metrics.

Low-Rank Adaptation (LoRA)~\cite{eHu21} is a technique to inject a number of trainable layers in the Transformer layers. The weights of those layers are then fine-tuned to the given task, while the remainder of the model's weights are frozen. These layers are also called adapters. We aim to understand the impact on performance with a model fine-tuned using LoRA, and providing multiple examples (from 0-shot to 8-shot) in prompt to the base and fine-tuned model.

\subsection{Datasets}

\subsubsection{}{Fine-tuning}
For fine-tuning LoRA adapters, the number of examples in train and val set size is $\approx27000$ and $\approx17000$ respectively.

\subsubsection{Evaluation Dataset}
\begin{table}[ht]
\centering
\caption{Evaluation dataset showing number of examples in each category.}
{\large
\begin{tabular}{@{}lccc@{}}
\toprule
\textbf{Name}           & \textbf{Car}         & \textbf{Service}      & \textbf{Time} \\
                        & \textbf{Details}     & \textbf{Request}     & \textbf{Negotiation} \\ \midrule
Auto-Rule-Based      & 4969                 & 2753                  & 2531 \\
%Rule-Based      &                      &                       & \\
Auto-Curated      & 235                  & 209                   & 230 \\ \bottomrule
%Curated           &                      &                       & \\    \bottomrule
\end{tabular}
}
\label{table:dataset_sizes}
\end{table}

Table \ref{table:dataset_sizes} shows the two evaluation datasets with the breakdown count for three conversation contexts, \emph{car details}, \emph{time negotiation}, and \emph{service request}, representing contexts where the user is expected to give information related to their vehicle, a preferred time slot, and service required for the vehicle, respectively. The two datasets are:

    \paragraph{Automotive-Rule-Based} This dataset consists of outputs generated by the rule-based pipeline. The metric used was the percentage of examples with matching predictions between the model and the rule-based pipeline. However, it is important to note that this dataset models the original pipeline behavior, which may not always be accurate.

    \paragraph{Automotive-Curated} This dataset contains human-validated labels and is more accurate for assessing the real-world performance of the models. To curate this dataset, first, we selected ambiguous examples with at least one overlapping entity. Then, each example was given to two annotators for labeling. Out of the $1193$ examples that were initially selected, annotators agreed on $674$ examples which created this dataset. 

\subsection{Models and Approaches}
In this section, we describe the two experiments with which we evaluate the performance of the LLM-based approach for entity disambiguation tasks. We use \texttt{Llama3.1-8B-Instruct}~\cite{aDubey24} model for the experiments in this paper.
\newline
    
    \subsubsection{Llama3.1-8B-Instruct Zero-shot and Multi-shot Prompting with manually curated examples} We evaluate the \texttt{Llama3.1-8B-Instruct} model’s performance by providing multiple examples, progressing from 1-shot to 8-shot. In this setup, the examples are manually curated and unique for each of the three contexts defined above. Multi-shot prompting allows the model to use contextual examples to better understand the input text, potentially improving performance with additional examples.

    \subsubsection{Llama3.1-8B-Instruct Multi-shot Prompting with Example-Based Retrieval} In our experiments with multi-shot prompting, we also implemeted an \emph{Example-Based Retrieval} pipeline~\cite{brown2020language,wang2023learning,black2021gpt, touvron2023llama} that selects examples based on the cosine similarity between the embeddings\footnote{\url{https://huggingface.co/sentence-transformers/all-MiniLM-L6-v2#all-minilm-l6-v2}} of the utterance and examples in the validation set.

    \subsubsection{Fine-tuned Llama3.1-8B-Instruct Zero-shot and Multi-shot Prompting} In this approach, we fine-tune \texttt{Llama3.1-8B-Instruct} model using LoRA, which makes it more computationally efficient while enhancing performance.
     
\section{Experimental Results}

\begin{table*}[ht!]
\centering
\caption{Overall Matching Percentage with a breakdown over three conversation contexts on \emph{Automotive-Rule-Based} dataset. Best numbers from LLM-based approach are highlighted in bold. All results are shown in percentage. For multi-shot prompting, \lq Example-Based $n$ shot' implies that the examples were obtained from the \emph{Example-Based Retrieval} pipeline, as opposed to manually curated examples.}
%\begin{adjustbox}{width=\textwidth}  % This ensures the table fits within the width of both columns
{\large
\begin{tabular}{@{}lccccc@{}}
\toprule
\textbf{Model}                                   & \textbf{Overall}   & \textbf{Car} & \textbf{Time} & \textbf{Service}      \\
                                                 & \textbf{Matches}    & \textbf{details}   & \textbf{negotiation}     & \textbf{request}  \\ \bottomrule
%Llama 3.1-8B-Instruct                            & 8.92                   & 15.38               & 4.15             & 1.79                \\ \midrule
%Llama 3.1-8B-Instruct + LoRA                     & 53.95                  & 54.60               & 49.39            & 53.69            \\ \midrule
Llama 3.1-8B-Instruct - 0 shot                       & 8.92                   & 15.38               & 4.15             & 1.79                \\ \midrule
Llama 3.1-8B-Instruct - 1 shot                   & 30.04                  & 33.95               & 34.26            & 19.11            \\ 
Llama 3.1-8B-Instruct - 2 shot                   & 46.63                  & 55.20               & 43.54            & 33.99            \\
Llama 3.1-8B-Instruct - 3 shot                   & 48.49                  & \textbf{55.77}               & 37.22            & 45.69            \\
Llama 3.1-8B-Instruct - 4 shot                   & 51.92                  & 55.18               & 43.11            & \textbf{54.16}           \\ 
Llama 3.1-8B-Instruct - 5 shot                   & 50.26                  & 55.12               & 42.47            & 48.67            \\
Llama 3.1-8B-Instruct - 6 shot                   & 50.76                  & 51.86               & 52.27            & 47.40            \\
Llama 3.1-8B-Instruct - 7 shot                   & \textbf{52.92}                  & 53.97               & 56.34            & 47.88            \\ 
Llama 3.1-8B-Instruct - 8 shot                   & 50.84                  & 47.01               & \textbf{57.29}            & 51.87            \\ \midrule
Llama 3.1-8B-Instruct - Example-Based 1 shot                   & 37.22                  & 45.80               & 22.87            & 34.94            \\ 
Llama 3.1-8B-Instruct - Example-Based 2 shot                   & 44.13                  & 54.28               & 28.55            & 40.14            \\ 
Llama 3.1-8B-Instruct - Example-Based 3 shot                   & 47.11                  & 56.93               & 32.54            & 42.79            \\ 
Llama 3.1-8B-Instruct - Example-Based 4 shot                   & 49.08                  & 56.91               & 36.06            & 46.93            \\ 
Llama 3.1-8B-Instruct - Example-Based 5 shot                   & 51.14                  & 58.74               & 38.11            & 49.40            \\ 
Llama 3.1-8B-Instruct - Example-Based 6 shot                   & 52.74                  & 59.73               & 39.38            & 52.41            \\ 
Llama 3.1-8B-Instruct - Example-Based 7 shot                   & 53.70                  & \textbf{60.27}               & 40.60            & 53.87            \\
Llama 3.1-8B-Instruct - Example-Based 8 shot                   & \textbf{53.73}                  & 60.03               & \textbf{40.84}            & \textbf{54.20}            \\ \midrule
Llama 3.1-8B-Instruct + LoRA - 0 shot                    & 53.95                  & 54.60               & 49.39            & 53.69            \\ \midrule
Llama 3.1-8B-Instruct + LoRA - 1 shot            & 75.72                  & 78.55               & 74.80            & 71.45   \\
Llama 3.1-8B-Instruct + LoRA - 2 shot            & \textbf{81.42}         & \textbf{82.29 }     & \textbf{79.19}   & \textbf{81.91 }  \\
Llama 3.1-8B-Instruct + LoRA - 3 shot            & 78.53                  & 80.72               & 71.72            & 80.82   \\
Llama 3.1-8B-Instruct + LoRA - 4 shot            & 79.41                  & 82.03               & 71.72            & 81.77   \\ 
Llama 3.1-8B-Instruct + LoRA - 5 shot            & 77.36                  & 79.71               & 70.93            & 79.00   \\ 
Llama 3.1-8B-Instruct + LoRA - 6 shot            & 80.16                  & 81.88               & 78.00            & 79.04   \\
Llama 3.1-8B-Instruct + LoRA - 7 shot            & 79.77                  & 80.70               & 78.87            & 78.93   \\
Llama 3.1-8B-Instruct + LoRA - 8 shot            & 76.70                  & 76.29               & 78.52            & 75.77   \\
\bottomrule

\end{tabular}
}
% \label{table:appendix_NLUpipelineeval}
\label{table:NLUpipelineeval}
\end{table*}

\begin{table*}[ht!]
\centering
\caption{Overall Matching Percentage with a breakdown over three conversation contexts on \emph{Automotive-Curated Dataset}. Best numbers from LLM-based approach are highlighted in bold. All results are shown in percentage. For multi-shot prompting, \lq Example-Based $n$ shot' implies that the examples were obtained from the \emph{Example-Based Retrieval} pipeline, as opposed to manually curated examples.}
%\begin{adjustbox}{width=\textwidth}  % This ensures the table fits within the width of both columns
{\large
\begin{tabular}{@{}lccccc@{}}
\toprule
\textbf{Model}                                   & \textbf{Overall}   & \textbf{Car} & \textbf{Time} & \textbf{Service}      \\
                                                 & \textbf{Matches}    & \textbf{details}   & \textbf{negotiation}     & \textbf{request}  \\ \bottomrule
NLU rule-based pipeline                               &    78.60   &   89.36   &   73.91   &   71.63               \\ \midrule
%Llama 3.1-8B-Instruct                       &   6.68	&   16.17   &   2.61	&   0.48                \\ \midrule
%Llama 3.1-8B-Instruct + LoRA                &   49.41   &	48.09   &   40.00	&   61.24               \\ \midrule
Llama 3.1-8B-Instruct - 0 shot                    &   6.68	&   16.17   &   2.61	&   0.48                \\ \midrule
Llama 3.1-8B-Instruct - 1 shot              &   53.11   &	35.32   &	69.57   &	55.02               \\ 
Llama 3.1-8B-Instruct - 2 shot              &   \textbf{65.28}   &	66.81   &	65.22   &	\textbf{63.64}               \\
Llama 3.1-8B-Instruct - 3 shot              &   60.83   &	\textbf{69.79}   &	58.70   &	53.11               \\
Llama 3.1-8B-Instruct - 4 shot              &   63.95   &	68.94   &	68.70   &	53.11               \\ 
Llama 3.1-8B-Instruct - 5 shot              &   62.61   &	65.11   &   \textbf{72.61}   &	48.80               \\
Llama 3.1-8B-Instruct - 6 shot              &   59.35   &	63.40   &   64.78   &   48.80               \\
Llama 3.1-8B-Instruct - 7 shot              &   60.39   &	63.40   &	68.26   &   48.33               \\
Llama 3.1-8B-Instruct - 8 shot              &   58.61   &	57.45   &   72.18   &	44.98               \\

\midrule
Llama 3.1-8B-Instruct - Example-Based 1 shot          & 51.63         & 51.91       & 55.22            & 47.36       \\ 
Llama 3.1-8B-Instruct - Example-Based 2 shot          & 60.68         & 61.28       & 66.09            & 54.07       \\ 
Llama 3.1-8B-Instruct - Example-Based 3 shot          & 64.39         & 68.94       & 70.87            & 52.15       \\ 
Llama 3.1-8B-Instruct - Example-Based 4 shot          & 67.36         & 65.96       & 78.26            & 56.94       \\ 
Llama 3.1-8B-Instruct - Example-Based 5 shot          & 65.88         & 66.38       & 75.22            & 55.02       \\ 
Llama 3.1-8B-Instruct - Example-Based 6 shot          & 68.25         & \textbf{68.94}       & 75.65            & 59.33       \\ 
Llama 3.1-8B-Instruct - Example-Based 7 shot          & 68.55         & 65.96       & 79.13            & 59.81       \\
Llama 3.1-8B-Instruct - Example-Based 8 shot          & \textbf{70.62}& 68.51       & \textbf{80.00}   & \textbf{62.68}            \\ \midrule
Llama 3.1-8B-Instruct + LoRA - 0 shot               &   49.41         &	48.09   &   40.00	&   61.24               \\  \midrule
Llama 3.1-8B-Instruct + LoRA - 1 shot       &   71.51   &   76.59   &   78.26   &   58.37               \\
Llama 3.1-8B-Instruct + LoRA - 2 shot       &   71.36   &	82.55   &	67.39   &	63.16               \\
Llama 3.1-8B-Instruct + LoRA - 3 shot       &   70.18   &	83.40   &	70.00   &   55.02               \\
Llama 3.1-8B-Instruct + LoRA - 4 shot       &   74.63   &   86.38   &	73.91   &	62.20               \\
Llama 3.1-8B-Instruct + LoRA - 5 shot       &   75.07   &	82.55   &	77.39   &	64.11               \\
Llama 3.1-8B-Instruct + LoRA - 6 shot       &   76.11   &	85.11   &	76.96   &	65.07               \\
Llama 3.1-8B-Instruct + LoRA - 7 shot       &   \textbf{78.93}      &	\textbf{87.23}   &	\textbf{82.18}   &	\textbf{66.03}               \\ 
Llama 3.1-8B-Instruct + LoRA - 8 shot       &   73.59	&   81.70   &	79.56   &	57.89               \\ \bottomrule
\end{tabular}
}
% \label{table:appendix_manualdataseteval}
\label{table:manualdataseteval}
\end{table*}

This section presents a comparison of the results from all approaches using the two evaluation datasets. The output for a given example is considered to be \emph{matching} when the predicted entities are exactly the same as the expected entities. We report results from multiple setups: using a base LLM and a fine-tuned LLM with LoRA, both in zero-shot and multi-shot settings~(see
Tables~\ref{table:NLUpipelineeval} and~\ref{table:manualdataseteval}).
% Refer to Tables~\ref{table:appendix_NLUpipelineeval} and~\ref{table:appendix_manualdataseteval} in the appendix for the full results.

\subsection{Automotive-Rule-Based Dataset Results}

Table~\ref{table:NLUpipelineeval} presents the matching percentage obtained by various approaches when evaluated on the \emph{Automotive-Rule-Based} dataset. We observe that the off-the-shelf LLM is not directly competitive for a domain specific task as the base model results in only 7\% matches when compared with the prediction of the NLU rule-based pipeline. However, after fine-tuning using the data from the specific domain, model gains a broader understanding of the entity conflict resolution task resulting in 53.95\% matches to the rule-based pipeline output.

When comparing the 0-shot LoRA fine-tuned model against multi-shot prompting applied to the base (not fine-tuned) model, we observe that the fine-tuned model achieves slightly higher accuracy overall, but the multi-shot approach outperforms the fine-tuned model on certain conversation contexts depending on the number of examples in the prompt. In particular, we observe a significant improvement for the \emph{time negotiation} context where the multi-shot LLM approach outperforms by 7.9\%~(in the 8-shot setting). We think this is because of LLM's inherent advantage with a broader knowledge context like \emph{time negotiation}.

We apply multi-shot prompting approach to the base model as well as the fine-tuned model to study if there are any concrete benefits that come from fine-tuning the model. Here, we observe that in general there is a progressive improvement in the multi-shot approach but there are non significant changes after 4 or 5 shot setting for both models. LoRA fine-tuned model in 2-shot setting gives the best result with an overall agreement of 81.42\% with the NLU rule-based pipeline.

In our data, we also observe that a lot of users are not always cooperative with the conversation agent. This means that it is possible that a user might give \emph{service request} information in an unrelated context. For such cases, multi-shot prompting where we give context specific examples is not very useful when applied to the base model. In contrast, multi-shot prompting applied to the fine-tuned model which has already captured the generic NLU pipeline behavior helps to learn better context specific behavior.

\subsection{Automotive-Curated Dataset Results}

Table~\ref{table:manualdataseteval} provides a comparison of the performance of the existing NLU rule-based pipeline against the proposed LLM solutions on the \emph{Automotive-Curated} dataset. %We observe a more steady progressive improvement for multi-shot settings compared to the NLU pipeline data results as shown in Figure~\ref{fig:trend}. This could potentially be an effect of less noise in the dataset. 

Once again, the base LLM model performs poorly overall (achieving only 6.68\% matches) with its multi-shot counterparts reaching up to 65.28\% overall accuracy. This outlines the importance of providing even few examples to help boost performance: adding 2 examples already yields almost 60\% increase in performance compared to 0-shot (see \texttt{Llama 3.1-8B-Instruct - 2 shot}).

Similarly, the LoRA fine-tuned model yields lower performance in its 0-shot setting, yet achieves the highest scores among LLM-based approaches, across all conversation contexts, in a multi-shot setting (78.93\% overall matches at $7$-shot setting). We have however noticed temporary performance drops when we add a new example. We think this is because the model needs time to recalibrate from the diversity introduced by the new example. Once it adjusts to the new pattern, its performance stabilizes and improves.

When comparing results across conversation contexts, we observe \emph{service request} benefits the most from fine-tuning as domain-specific knowledge is of high importance for such context (almost 60\% increase from the base to the fine-tuned model in 0-shot setting), whereas for the other two contexts, providing the base model (not fine-tuned) with examples outperforms the 0-shot LoRA fine-tuned variant in all multi-shot settings.

In comparison to the existing rule-based NLU pipeline, the best LLM-based approach is overall competitive to the existing approach's matching accuracy of 78.60\%. However, when we breakdown the results across conversation contexts for the best LLM multi-shot setting, we observe that the LLM-based approach outperforms existing NLU pipeline for \emph{time negotiation} context requests by 8.27\%. Contrary to these results, for \emph{service request} and \emph{car details}, NLU outperforms by 5.6\%, and 2.13\% respectively. These results are aligned with our hypothesis that LLMs perform better at generic contexts like \emph{time negotiation} compared to a domain specific contexts like \emph{service request}.

In the next section, we present an analysis on the findings from the experimental results.

\begin{figure}[ht]
    \centering
    \includegraphics[width=0.49\textwidth]{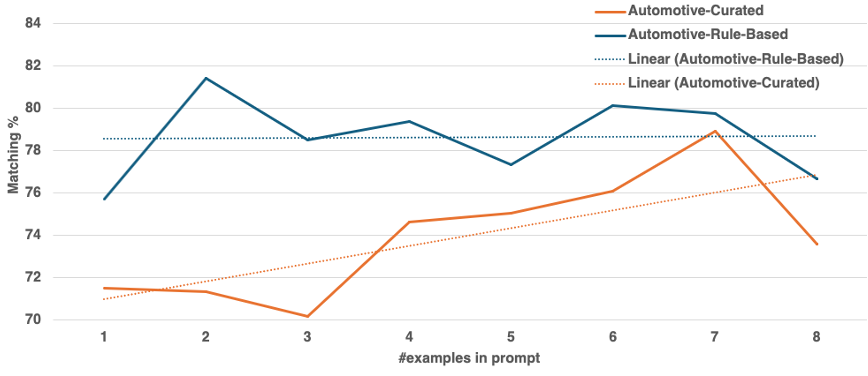} % Replace with actual figure
    \caption{Difference in trends of improvement between both datasets when multi-shot prompting with manually curated examples is applied to fine-tuned model.}
    \label{fig:trend}
\end{figure}

\begin{figure}[ht!]
    \centering
    % First Subfigure
    \begin{subfigure}{0.49\textwidth}
        \centering
        \includegraphics[width=\textwidth]{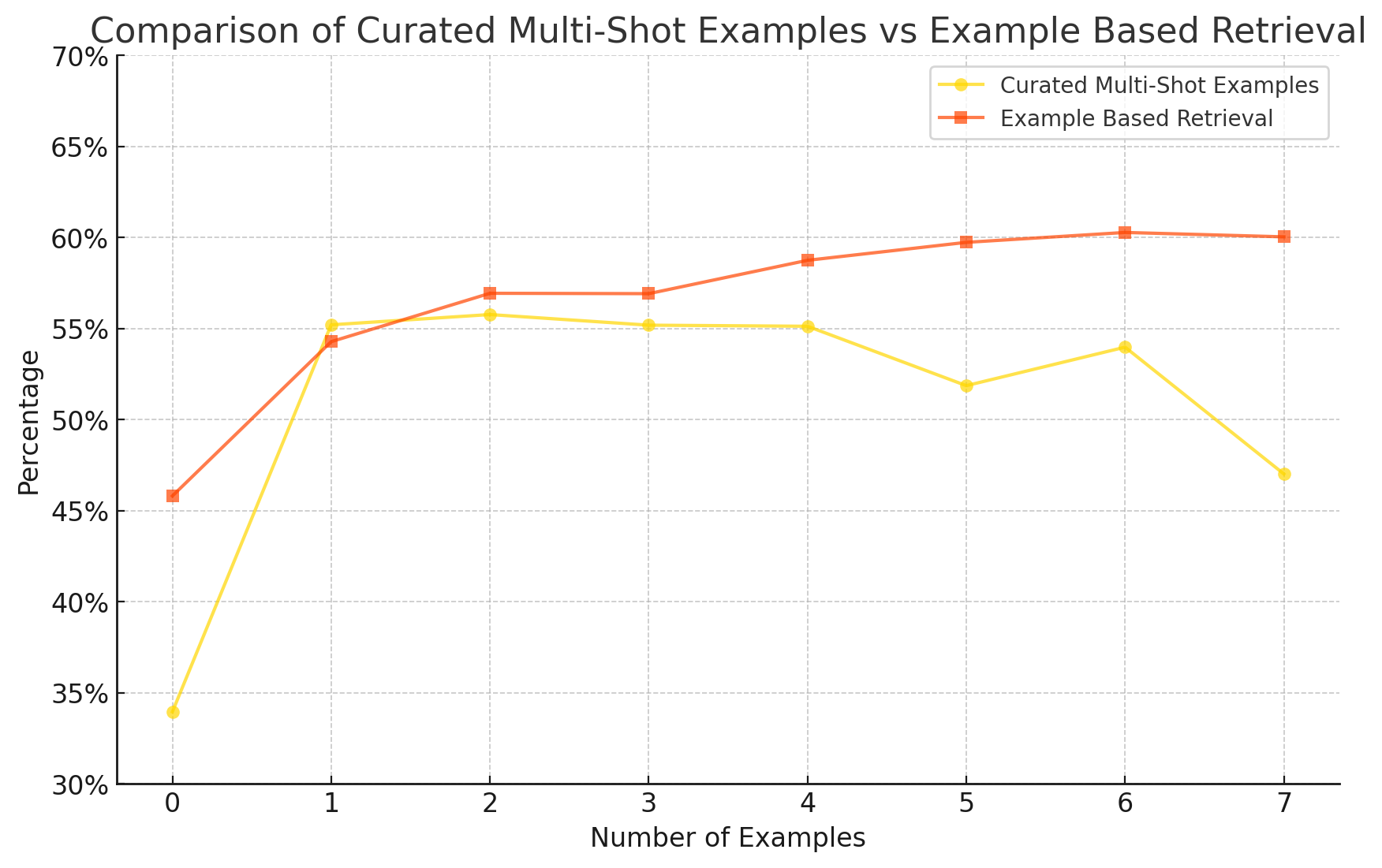} % Replace with actual figure
        \caption{\emph{car details} context.}
        \label{fig:examplebasedvs}
    \end{subfigure}
    % Second Subfigure
    \begin{subfigure}{0.49\textwidth}
        \centering
        \includegraphics[width=\textwidth]{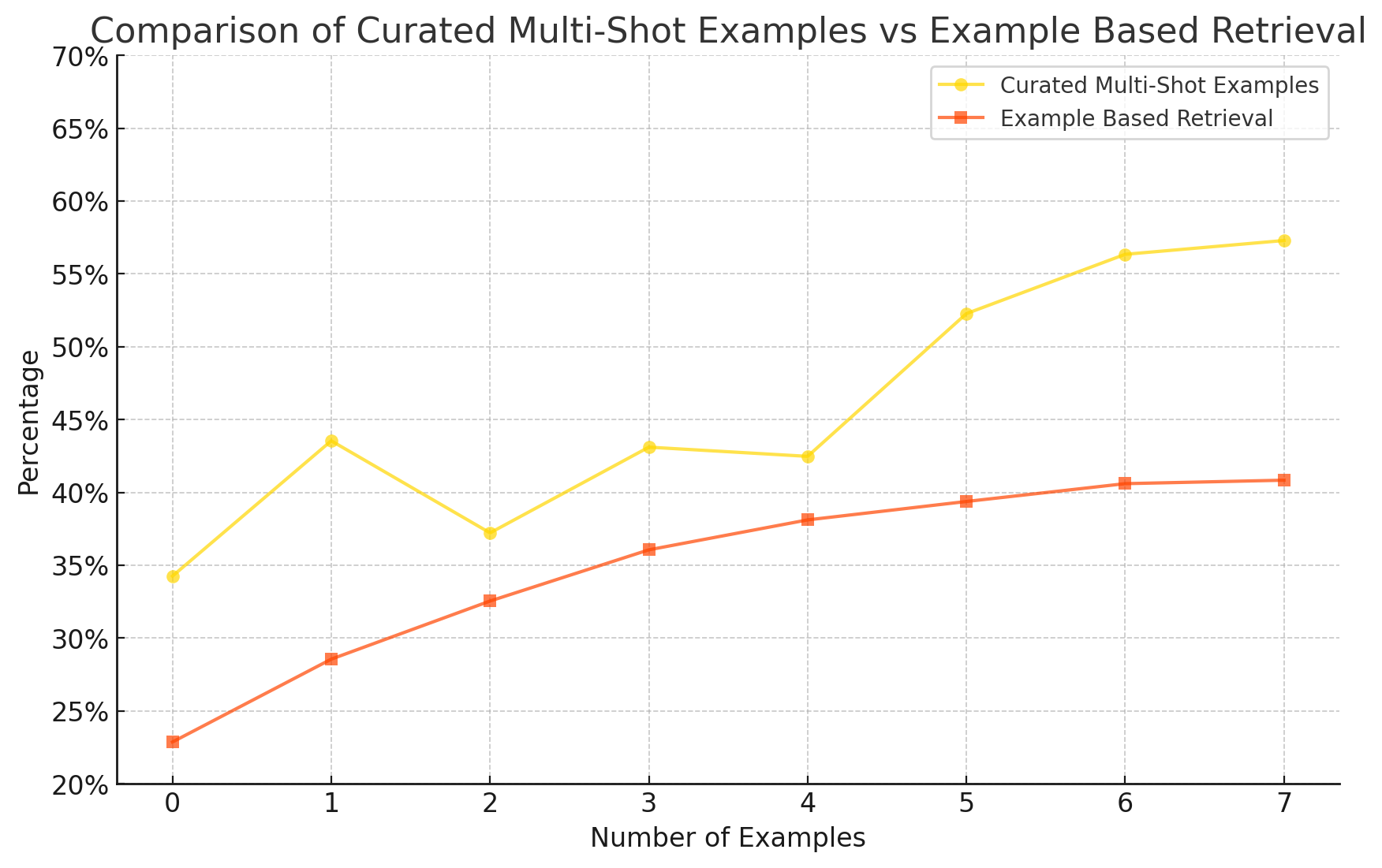} % Replace with actual figure
        \caption{\emph{time negotiation} context.}
        \label{fig:examplesdsbasedvs}
    \end{subfigure}
    % Third Subfigure
    \begin{subfigure}{0.49\textwidth}
        \centering
        \includegraphics[width=\textwidth]{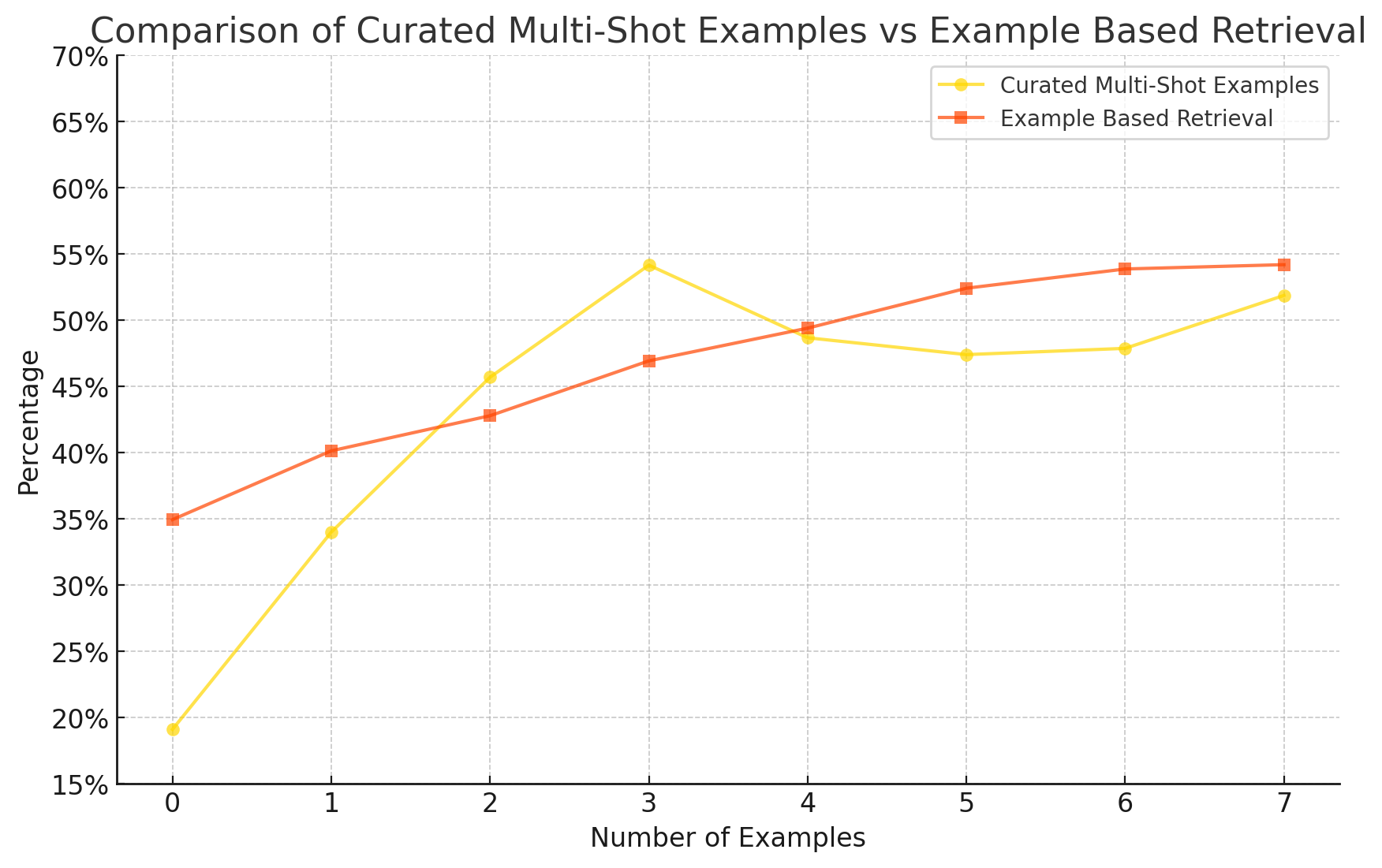} % Replace with actual figure
        \caption{\emph{service request} context.}
        \label{fig:examplesdsbasedvs} 
    \end{subfigure}
    \caption{Comparison of trends in improvement when examples in multi-shot prompting are manually curated versus when they are obtained from \emph{Example-Based Retrieval} pipeline. Results shown above are obtained using \emph{Llama 3.1-8B-Instruct model} on \emph{Automotive-Rule Based} dataset. a) \emph{car details} context b) \emph{time negotiation} context c) \emph{service request} context.}
    \label{fig:examplevscurated}
\end{figure}

\section{Comparison and Analysis}
\subsection{Performance on Domain specific cases} Table~\ref{table:manualdataseteval} shows that the best LLM-based approach significantly outperforms the rule-based approach for generic conversational contexts like \emph{time negotiation}, offers a comparative performance to a less generic context like \emph{car details}, and underperforms in domain specific context like \emph{service request}. We observe a trend where the performance of LLM drops as the domain-specificity increases. For example, in the \emph{time negotiation} context, given the utterance \emph{`december fifteen two thousand and thirteen'}, and two possible entity types: \emph{car year} and \emph{time}, LLM correctly leverages the context to remove the \emph{car year} entity, whereas the rule-based approach fails at doing the same because of higher priority for \emph{car year}. In another example from \emph{car details} context, \emph{`two thousand fifteen ram fifteen hundred'}, the LLM approach retains the \emph{car year}, and \emph{make} (`ram'), but fails to recognize `ram 1500' as a valid \emph{car\_model} entity. Rule-based method outperforms here because it prioritizes maximum span coverage and removes \emph{make}.

\subsection{Difference in progressive improvement between both datasets} Figure~\ref{fig:trend} shows the difference in trends of improvement across both datasets when multi-shot prompting is applied to the fine tuned model (best case setting). We see a clear increasing trend for \emph{Automotive-Curated} dataset as compared to \emph{Automotive-Rule-Based} which could be attributed to the noise in the latter.

\subsection{Performance gains from fine-tuning} For a generic context like \emph{time negotiation}, multi-shot applied to the base model is also quite close to NLU pipeline. However, for domain specific cases, major performance gains come from fine-tuning.

\subsection{Gains from \emph{Example-Based Retrieval} pipeline}
In Figure~\ref{fig:examplevscurated}, we compare the improvement trends in a multi-shot prompting setup when using an \emph{Example-Based Retrieval} pipeline versus manually curated examples. The \emph{Example-Based Retrieval} retrieval pipeline consistently demonstrates a steady upward trend across all conversational contexts, whereas the curated examples exhibit fluctuations depending on the newly added example and its quality. In terms of performance, the \emph{Example-Based Retrieval} pipeline outperforms curated examples in the \emph{car details} context, likely because similar car makes and models share common issues, making relevant examples more accessible in the embedding space. However, in the \emph{time negotiation} context, semantically similar examples in the embedding space do not necessarily provide equivalent information.

\section{Conclusion}
In this work, we evaluated the performance of the~\texttt{Llama3.1-8B-Instruct} model using both multi-shot prompting and fine-tuning approaches, for the task of entity conflict resolution in a pipeline-based NLU architecture. Our experiments evaluated the performance of multi-shot prompting and LoRA fine-tuning in multiple settings across two datasets: \emph{Automotive-Rule-Based}, derived from the existing NLU pipeline and \emph{Automotive-Curated}, obtained through manual annotations providing a reliable source of correctness. The results demonstrated that while the multi-shot prompting approach on the base model provides flexibility in behavior adjustment with minimal training effort, its performance is lower in comparison to the rule-based NLU pipeline, especially for domain specific contexts.

Ultimately, the best LLM-based approach appears to be a hybrid strategy, where multiple examples are provided in prompt to the LoRA fine-tuned~\texttt{Llama3.1-8B-Instruct} model. This approach offers significant gains in accuracy reaching a comparable performance with rule-based system for two out of three conversation contexts. This balanced approach is a good solution to quickly adapt models in the short term while achieving sustained performance improvements through fine-tuning. Though increased latency is a potential trade-off, replacing rule-based systems with LLMs offers significant advantages in terms of flexibility and scalability across diverse and complex tasks.

For future work, we believe that there is a large margin of improvement with a sophisticated \emph{Example-Based Retrieval} strategy~\cite{liu2021makes,rubin2021learning} that also focuses on diversity of the newly added examples.
% For future work, we believe that there is a large margin of improvement with example based retrieval strategies~\cite{wang2023learning}. Further, we would also like to expand the scope of this work to other conversational contexts in the NLU pipeline.

\Large{
\bibliography{custom.bib}
}
% \appendix

% \section{Results}
% The complete data from Tables~\ref{table:NLUpipelineeval} and~\ref{table:manualdataseteval} from the main text is given below:

% \section{Example Appendix}
% \label{sec:appendix}

% This is an appendix.

\end{document}